\documentclass[aps,prb,twocolumn]{revtex4}
\usepackage{epsfig}
\usepackage{bm}
\usepackage{wasysym}
\newcommand{\spn}{{\mathrm{Sp}(N)}}
\begin{document}

\title{Flux expulsion and greedy bosons: frustrated magnets at
large $N$}

\author{O. Tchernyshyov} 
\affiliation{Department of Physics and Astronomy, Johns Hopkins University,
Baltimore, Maryland, 21218}
\author{R. Moessner} 
\affiliation{Laboratoire de Physique Th\'eorique de l'Ecole Normale
Sup\'erieure, CNRS-UMR8549, Paris, France}
\author{S. L. Sondhi} 
\affiliation{Department of Physics, Princeton University,
Princeton, New Jersey 08544} 

\date{August 23, 2004}

\begin{abstract}
  We investigate the $\spn$ mean-field theory for frustrated quantum
  magnets. First, we establish some general properties of its
  solutions; in particular, for small spin we propose simple rules for
  determining the saddle points of optimal energy. We then apply these
  insights to the pyrochlore lattice. For spins on a single
  tetrahedron, we demonstrate a continuous ground state degeneracy for
  any value of the spin length. For the full pyrochlore lattice,
  this degeneracy translates to a large
  number of near-degenerate potential saddle points.
  Remarkably, it is impossible to construct a
  saddle point with the full symmetry of the Hamiltonian---at large
  $N$, the pyrochlore magnet {\em cannot} be a spin
  liquid. Nonetheless, for realistic finite values of $N$, tunnelling
  between the nearly degenerate saddle points could restore the full
  symmetry of the Hamiltonian.
\end{abstract}

\maketitle

{\bf Introduction.} 
The behavior of quantum magnets with strong frustration\cite{greedan} 
is one of the central open questions in the study of modern magnetism. The
most celebrated members of this class of problems are the
nearest-neighbor Heisenberg antiferromagnets on the kagome and
pyrochlore lattices and the challenge is to work out their 
phases at varying values of the spin $S$ and temperature $T$. 
At large $S$, semiclassical computations become feasible if not 
straightforward.\cite{shender} Small spins have typically been treated by 
entirely separate methods, and for the case of the
pyrochlore lattice these have involved starting from a symmetry-breaking 
decomposition of the lattice.\cite{ml2003}

As a complement to this work, we employ 
a large-$N$ method that can be applied over the full range of $S$
and, particularly for the work described in this paper,
allows us to keep all symmetries in the formalism at small $S$.
Specifically, we consider the enlargement of the 
SU(2) $\equiv$ Sp(1)
symmetry group of the Heisenberg model to Sp($N$).  In the limit 
$N \to \infty$, the problem reduces to a mean-field theory for Schwinger 
bosons.\cite{arovas88,read90,sachdev92}  
The mean-field theory can be 
improved by evaluating $1/N$ corrections.
This method has been applied to the kagome lattice\cite{sachdev92prb} 
and found to predict a robust
selection of a magnetically ordered state at large $S$
melting into a disordered spin-liquid state at small $S$. 

We first establish some general properties of 
mean-field ground states of $\spn$ magnets on regular lattices
at small values of spin.
The main variables of the mean-field theory are complex numbers $Q_{ij}$
defined on links $(ij)$ of the lattice and representing probability
amplitudes of finding a valence-bond singlet.  We derive a perturbative 
loop expansion, wherein spin length is used as a small parameter.
To the lowest nontrivial order, valence bonds are found {\em exclusively} 
on links with largest exchange coupling.  At higher orders, we find 
that a U(1) flux, constructed from phases of the valence-bond amplitudes,
tends to be expelled.  
These observations systematize previous results obtained for
several quantum antiferromagnets in the large-$N$ 
framework.\cite{sachdev92prb} 
 
We then discuss the case of the pyrochlore antiferromagnet where
we report three principal results: (i) For
a single tetrahedron, the two-parameter degeneracy 
of classical spins persists for {\em any} spin length despite the
quantum effects encoded in the $\spn$ computation! 
(ii) All of these solutions break a symmetry---spatial or time 
reversal---and exhibit bond or chiral order. (iii)  Embedding such states
in the pyrochlore lattice results in a vast number of solutions whose
energies are very close to one another at small spin. This behavior is 
fundamentally different from the triangular and kagome lattices, where 
the principle of flux
expulsion alone was sufficient to fix the ground states.
The import of (ii) and (iii) is that, if the pyrochlore lattice is indeed a
spin liquid at $N = 1$, this will necessarily require tunnelling
between saddle points.  Such a scenario is not at all unlikely.

{\bf $\spn$ Hamiltonian.}
The Hamiltonian of an $\spn$ antiferromagnet may be written 
as\cite{sachdev92prb}
\begin{equation}
H = -(1/2N)\sum_{\langle ij \rangle} J_{ij}
{\mathcal A}^\dagger_{ij}{\mathcal A}_{ij}, 
\hskip 3mm
{\mathcal A}_{ij} = \varepsilon_{\alpha\beta} b_{i\alpha} b_{j\beta}.
\label{eq-SpN-H}
\end{equation}
The antisymmetric tensor $\varepsilon_{\alpha\beta}$ is a block-diagonal 
$2N\!\times\!2N$ matrix
\begin{equation}
\varepsilon = 
\left(
\begin{array}{ccc}
i\sigma_y & 0 & \ldots\\
0 & i\sigma_y & \ldots\\
\vdots & \vdots 
\end{array}
\right)
\label{eq-eps-def}
\end{equation}
and $\sigma_y$ is the $2\times2$ Pauli matrix.  
Here, $b_{i\alpha}$ is the annihilation operator of a boson of species
$\alpha$ at site $i$.  
For $N=1$, Eq.~(\ref{eq-SpN-H}) reduces to the SU(2)
Heisenberg Hamiltonian written in terms of Schwinger bosons, 
related to spin operators via $S_i^a =
b^\dagger_{i\alpha}
\sigma^a_{\alpha\beta} b_{i\beta}/2$ (a sum over doubly repeated 
flavor indices is implied); the number of bosons determines spin 
length: $b^\dagger_{i\alpha} b_{i\alpha} = 2S$.  In the large-$N$ 
generalization,
spin length is related to the number of bosons 
{\em per flavor}\cite{sachdev92prb}
$\kappa\equiv b_{i\alpha}^\dagger b_{i\alpha}/N$.

{\bf Mean-field approximation}
The $\mathrm{Sp}(N)$ mean-field
equations,\cite{sachdev92} which become exact for 
$N \to \infty$,
involve a decoupling of the quartic terms in Eq.~(\ref{eq-SpN-H}) 
with the aid of the link variables
$
\langle b_{i\alpha} b_{j\beta} \rangle 
= Q_{ij} \varepsilon_{\alpha\beta}/2$:
\begin{eqnarray}
H_{\rm MF} &=& (J_{ij}/2) \sum_{\langle ij \rangle} 
\big( N |Q_{ij}|^2 
- Q_{ij} \varepsilon_{\alpha\beta} b^\dagger_{i\alpha} b^\dagger_{j\beta}
\nonumber
\\
&-& Q_{ij}^* \varepsilon_{\alpha\beta} b_{i\alpha} b_{j\beta} \big)
+ \sum_{i} \lambda_i (b^\dagger_{i\alpha} b_{i\alpha} - \kappa N).
\label{eq-HMF}
\end{eqnarray}
The chemical potential $\lambda_i$ keeps the average number of bosons 
fixed at $\kappa N$ on every site.  The mean-field equations are
$\partial \langle H_{\rm MF} \rangle / \partial \lambda_i = 0$ 
(constraints on the boson numbers) and 
$\partial \langle H_{\rm MF} \rangle / \partial Q_{ij} = 0$ 
(minimization of energy).

The mean-field Hamiltonian (\ref{eq-HMF}) is a sum of $N$ identical
copies, each containing two flavors only (up and down in the Schwinger-boson 
language).  The energy of each copy is ${\mathcal O}(1)$ in
$N$.  Therefore, two different vacua will have an energy difference
${\mathcal O}(N)$, i.e., well separated in the limit
$N\to\infty$ considered here.  The low-lying excitations are
$S=1/2$ bosons (spinons) whose energy is ${\mathcal O}(1)$.

For a total of ${\mathcal N}$ sites, this Hamiltonian can, for each
pair of flavors, be written in terms of ${\mathcal N}$-component row
vectors $b^\dagger_{i\uparrow}$ and $b_{i\downarrow}$, column vectors
$b_{i\uparrow}$ and $b^\dagger_{i\downarrow}$, and ${\mathcal
N}\!\times\!{\mathcal N}$ matrices $P_{ij} = J_{ij}Q_{ij}/2$ and
$\Lambda_{ij} = \lambda_i\delta_{ij}$. 
A Bogoliubov transformation diagonalizes the
part quadratic in bosons and yields a diagonal matrix of
eigenfrequencies $\Omega$ for the bosonic spinons giving the energy
per flavor
\begin{eqnarray}
\langle H_{\rm MF} \rangle/N = 
 {\rm Tr}\left[ PQ^\dagger/2 - (\kappa+1) \Lambda  + \Omega \right]
\label{eq-E-per-flavor}
\end{eqnarray}

{\bf Uniform $\lambda$}.  
In the remainder of this paper we consider the mean-field theory with
two restrictions.  Firstly, we 
have tacitly assumed the absence of a condensate of bosons, $\langle
b_{i\alpha}\rangle\equiv 0$.  
(This translates into the lack of magnetic order, 
$\langle{\bf S}_i\rangle = 0$, for SU(2) spins.)  This regime, dominated by 
quantum fluctuations, always exists for small values of
spin length $\kappa$.\cite{arovas88,sachdev92prb}  In addition, we will 
only consider states with uniform chemical potential, so that 
$\Lambda = \lambda {\bm 1}$, where $\bm 1$ is the 
${\mathcal N} \times {\mathcal N}$ unit matrix.  
This simplifies calculations as matrices $\Lambda$ and $P$ commute.

With said restrictions, the boson spectrum $\Omega_n$ can be obtained 
from the eigenvalues $\nu_n^2$ of the matrix $PP^\dagger$:
\begin{equation}
\Omega_n = \sqrt{\lambda^2 - \nu^2_n},
\hskip 5mm
\det{(PP^\dagger - \nu^2_n {\bm 1})} = 0.
\label{eq-nu-def}
\end{equation}
The expectation value of the Hamiltonian is then
\[
\langle H_{\rm MF} \rangle/N 
= {\rm Tr}\left[ PQ^\dagger/2 + \sqrt{\lambda^2 {\bm 1} -PP^\dagger} 
-\lambda(1+\kappa){\bm 1}\right].
\]
The boson-number constraint gives the condition
\begin{equation}
1+\kappa = 
{\rm Tr} \left[ ({\bm 1} - PP^\dagger/\lambda^2)^{-1/2} \right] 
/ {\rm Tr} \, {\bm 1}.
\label{eq-constraint-matrix}
\end{equation}
A scaling transformation $Q_{ij} \to a Q_{ij},\  
\lambda \to a \lambda$
does not affect the constraint equation (\ref{eq-constraint-matrix}).
Minimization of the energy determines the optimal scale $a$
and yields
\begin{equation}
\frac{\langle H \rangle}{N} = 
- \frac{
\left\{ {\rm Tr} \left[ 
\left( PP^\dagger/\lambda^2 \right) 
\left( {\bm 1} - PP^\dagger/\lambda^2 \right)^{-1/2}
\right] \right\}^2
}{ 
2\, {\rm Tr} \left( PQ^\dagger/\lambda^2 \right) 
}.
\label{eq-E-matrix}
\end{equation}

Further minimization of the vacuum energy (\ref{eq-E-matrix}) is done
by varying the relative strengths and phases of the link variables
$Q_{ij}$, subject to constraint (\ref{eq-constraint-matrix}).  Our
{\em ad-hoc} assumption of a uniform chemical potential restricts the
choice of trial states in that all sites must be equivalent.

{\bf Loop expansion at small $\kappa$}.
The structure of the mean-field equations (\ref{eq-constraint-matrix}) 
and (\ref{eq-E-matrix}) suggests a solution by expansion in
powers of $PP^\dagger/\lambda^2$.  Taking the trace makes it a 
loop expansion: a generic term of the Taylor-expanded right-hand side of
Eq.~(\ref{eq-constraint-matrix}) has the form
\begin{equation}
{\rm Tr} \left( PP^\dagger\right)^n 
= \sum_{a \ldots z} P_{a b} (-P_{b c}^*)
\ldots P_{y z} (-P_{z a}^*)
\equiv \Xi e^{i\Phi},
\label{eq:fluxdef}
\end{equation}
where we have accounted for antisymmetry, $P_{c b} = -P_{b c}$.
Expressions for $\kappa$ and $E = \langle H \rangle$ involve sums over
all possible closed loops $abc\ldots za$ of even length.  The U(1) flux
$\Phi$ defined in Eq.~(\ref{eq:fluxdef}) will play an important role 
below.\cite{fn-flux}

Convergence is particularly good when spinons have a short correlation
length, as is the case for small $\kappa$.  In this limit,
the physics is determined by the shortest 
loops.  Formally, the loop expansion is a series in powers 
of $\kappa$.
In the following paragraphs, we develop the loop expansion and
demonstrate that it leads to simple organizing principles for the
behavior of $\spn$ antiferromagnets in the quantum limit of small
$\kappa$. A variant of this strategy at high temperatures has
been described previously.\cite{ts02}

{\bf Shortest loops, greedy bosons}.
The lowest order in $PP^\dagger/\lambda^2$ yields the energy per site,
per flavor:
\begin{equation}
\frac{\langle H \rangle} {{\mathcal N}N} = - \kappa \,
\frac{
{\rm Tr} \left( PP^\dagger \right)
}{
{\rm Tr} \left( PQ^\dagger \right)
}
= - \frac{\kappa}{2} \, 
\frac{
\sum_{\langle ij \rangle} J_{ij}^2 |Q_{ij}|^2
}{
\sum_{\langle ij \rangle} J_{ij} |Q_{ij}|^2
}.
\end{equation}
To leading order (``loops'' of length 2), the energy depends on the
absolute values, but not the phases of the link variables $Q_{ij}$.
Minimization can be easily done if one interprets $J_{ij}|Q_{ij}|^2$
as a probability distribution.  The energy is then simply the
expectation value of $-\kappa J_{ij}/2$.  An optimal probability
distribution then will have zero probabilities $J_{ij}|Q_{ij}|^2$ for
all links {\em except} those with the largest $J_{ij}$.  For example,
on a square lattice with first and second-neighbor couplings $J_1,
J_2>0$, the second-neighbor bonds will vanish, $Q_2=0$, if $J_1>J_2$;
similarly, $Q_1=0$ if $J_1<J_2$ in this approximation.  Hence
\begin{quote}
{\sc Theorem} (greedy bosons): in the limit of small $\kappa$, bosons
form bonds $Q_{ij} \neq 0$ on the links with the largest $J_{ij}$ only.
\end{quote}

For small but finite $\kappa$, there will be three phases: (i) $Q_2 = 0$ for
$J_2/J_1$ below a critical value $(J_2/J_1)_{c1} < 1$; (ii) $Q_1 = 0$ for
$J_2/J_1$ above another critical value $(J_2/J_1)_{c2} > 1$; and (iii)
coexisting $Q_1 \neq 0$ and $Q_2 \neq 0$ for intermediate values of $J_2/J_1$.

{\bf Longer loops, flux expulsion}.
The terms of order $(PP^\dagger)^2$ represent loops
containing up to 4 links.  For a single (e.g., nearest-neighbor) nonzero 
exchange constant $J$, $PQ^\dagger=2PP^\dagger/J$ and the energy, to this 
order, is
\begin{equation}
\frac{\langle H \rangle} {{\mathcal N}N} 
= - \frac{\kappa J}{2}
- \frac{\kappa^2 J}{4}\, 
\frac{{\rm Tr}\! \left[\left( PP^\dagger \right)^2\right] {\rm Tr}\, 1}
{\left[{\rm Tr} \left( PP^\dagger \right)\right]^2}.
\end{equation} 
For a fixed ``norm'' ${\rm Tr} (PP^\dagger) = \sum_{\langle ij
\rangle} J_{ij}^2 |Q_{ij}|^2$, lower energy means larger ${\rm Tr}
(PP^\dagger PP^\dagger)$.  This can be achieved by tuning the {\em
phases} of link fields $Q_{ij}$, as the contributions to ${\rm Tr}
(PP^\dagger PP^\dagger)$ from loop $abcda$ and its reverse are 
$P_{ab} P_{bc}^* P_{cd} P_{da}^* 
+ \mbox{C.c.}
= 2\Xi \cos{\Phi}$,
cf.\ Eq.~(\ref{eq:fluxdef}).
Clearly, for fixed magnitudes $|P_{ij}|$, the trace is 
maximized---and energy is minimized---when the flux $\Phi$ vanishes.  

This establishes the principle of flux expulsion for the shortest
nontrivial loops (length 4). As the loop expansion at small
$\kappa$ is organised by loop length, this principle provides the
correct ground state as the (ideally, only) one in which all loops up
to a certain length contain no flux. For example, it uniquely
determines the $\spn$ ground states observed for the uniform
triangular and kagome antiferromagnets.\cite{sachdev92prb}

More generally, we can formulate a conjecture on the behavior of
longer loops. It provides {\em all} gauge-invariant information about
the phases of link variables $Q_{ij}$.  
\begin{quote}
{\sc Conjecture} (flux expulsion): In the ground state, 
the flux $\Phi$ is zero through all closed loops of even 
length, provided such a fluxless state is possible.
\end{quote}
Keep in mind that the {\em tendency to expel} flux does not always 
guarantee the actual {\em absence} of flux.  If the lattice is
not bipartite, fluxes may be frustrated and will not be expelled from
every loop. This happens already for the triangular and kagome
cases and for the latter this makes the selection more delicate 
than believed previously.\cite{tmslong} It happens with a
vengeance on the pyrochlore lattice.

{\bf Pyrochlore I: the single tetrahedron}.
The shortest loop (of even length) on the pyrochlore lattice contains four
links and is confined to a single tetrahedron.  There are, in fact, three 
such loops on every tetrahedron and it can be verified that 
the sum of their fluxes equals $\pi$ (unless some link amplitudes vanish). 
The fluxes are thus frustrated and cannot be expelled from all three loops.
Unlike in all previously studied systems, the principle of flux expulsion 
does not point to a unique ground state.  

\begin{figure}
\centerline{\includegraphics[width=0.5\columnwidth]{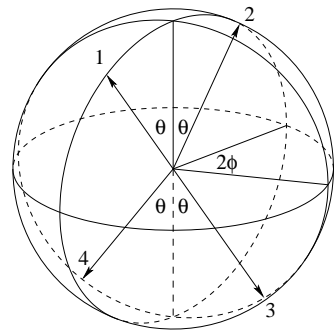}}
\caption{The two parameter family of ground states of a single tetrahedron. 
The four spins combine to give zero total spin.}
\label{fig-sphere}
\end{figure}

In fact, for the single tetrahedron, we find a two-parameter family of
ground states, all with exactly the same mean-field energy $E/(JN\mathcal
N) = -\kappa(\kappa+1)/2$. These have link variables
\begin{eqnarray}
Q_{12} = Q_{34} &=& \sqrt{\kappa(\kappa+1)} \sin{\theta}, 
\nonumber\\
Q_{13} = Q_{24} &=& \sqrt{\kappa(\kappa+1)} 
(\cos{\phi} - i \cos{\theta} \sin{\phi}), 
\label{eq-Q-tetra-exact}\\
Q_{14} = Q_{23} &=& \sqrt{\kappa(\kappa+1)} 
(\cos{\theta} \cos{\phi} - i \sin{\phi}). 
\nonumber
\end{eqnarray}
This is quite remarkable as one of the charms of $\spn$ is its capacity
to yield {\em unique} quantum disordered states at small $\kappa$---indeed, 
this is the first counterexample!
Remarkably, there is a mapping between these ground states and those of
{\em classical} Heisenberg spins on a tetrahedron, which can be
constructed by (a) parametrizing the ground states as shown in
Fig.~\ref{fig-sphere}, (b) representing the spins ${\bf S}_i$ by 
two-component spinors 
$\psi_{i\alpha}$ and (c)
translating the spinors into link variables 
$Q_{ij} \propto \varepsilon_{\alpha\beta} \psi_{i\alpha} \psi_{j\beta}$.

{\bf Possible orders}.
All of these mean-field ground states violate a symmetry of the
Hamiltonian (\ref{eq-SpN-H}): a point-group symmetry, time reversal,
or both.
These symmetry breakings are best illustrated by the sets of states
which break only a single symmetry.

\begin{figure}
\centerline{\includegraphics[width=\columnwidth]{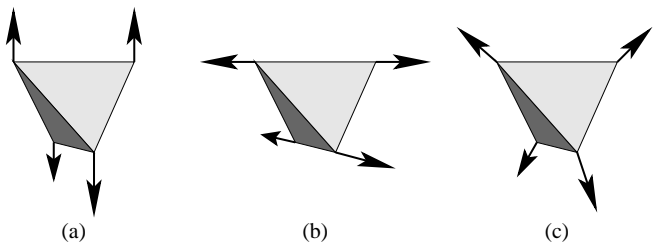}}
\caption{Trial ground states of the single tetrahedron: (a) collinear, 
(b) coplanar, (c) anticollinear.}
\label{fig-pyrochlore-states}
\end{figure}

First, breaking the symmetry group of the tetrahedron $T_d$, are 
three {\em bond-ordered} states 
with maximally inhomogeneous bond amplitudes, e.g. $Q_{12} = Q_{34} = 0$ and
$ Q_{13} = Q_{24} = Q_{14} = Q_{23} \neq 0$.  These we call the
``collinear'' states, as their classical counterparts have collinear
spins [Fig.~\ref{fig-pyrochlore-states}(a)].  The flux through  
loop 13241 vanishes; the other two fluxes are ill-defined.
The valence-bond order parameter characterizing the broken symmetry 
is described in Refs.~\onlinecite{hbb} and \onlinecite{tch02}.

Second are states which leave the spatial symmetry
intact but break the time-reversal symmtery.  These have $\theta =
\frac{1}{2}\arccos{(-1/3)}$, $\phi = \pm\pi/4$.  They distribute the
flux $\pi$ equally between the three loops, each receiving 
either $+\pi/3$ or $-\pi/3$.  The classical analogs of the two states have 
spins pointing at equal angles of $\arccos{(-1/3)} \approx 109^\circ$
to each other [Fig.~\ref{fig-pyrochlore-states}(c)]. 
The order parameter is spin {\em chirality} 
$\chi = \langle {\bf S}_a \cdot ({\bf S}_b \times {\bf S}_c)\rangle$,
where $abc$ is an oriented face of the tetrahedron.\cite{tsune,tch02}

{\bf Pyrochlore II: the full lattice}. 
The pyrochlore lattice is a network of corner-sharing
tetrahedra.  In the loop expansion up to ${\mathcal O}(\kappa^2)$,  
the system behaves as if it were made up of disjoint simplices: 
the energy is minimized as long as each tetrahedron is in any of 
its ground states (\ref{eq-Q-tetra-exact}).  This extremely large
degeneracy will be lifted, at least partially, at $\mathcal 
O(\kappa^3)$, which includes loops enclosing the hexagons of
the pyrochlore lattice.  In the spirit of degenerate perturbation theory,
we must minimize the terms $\mathcal O(\kappa^3)$ among all such possible
states.  This is a problem of considerable complexity.

On a first approach we have minimized the energy over a 
restricted set of trial states in 
which all tetrahedra are in the same state (\ref{eq-Q-tetra-exact}). 
Evaluation of the energy for a finite
cluster shows that, in accordance with our earlier conjecture, the 
energy reaches a minimum when the hexagons enclose zero flux; this choice
is similar to Sachdev's $Q_1=-Q_2$ state on the
kagome.\cite{sachdev92prb}

We obtain the following energies per site for the
ground states derived from classical spin states shown in
Fig.~\ref{fig-pyrochlore-states}:
\begin{equation}
\frac{E}{JN\mathcal N} = 
-\frac{\kappa}{2} -\frac{\kappa^2}{2}
- \left\{ \begin{array}{ll}
0.18750 \kappa^3 & \mbox{Fig.~\ref{fig-pyrochlore-states}(a)},\\
0.19922 \kappa^3 & \mbox{Fig.~\ref{fig-pyrochlore-states}(b)},\\
0.20139 \kappa^3 & \mbox{Fig.~\ref{fig-pyrochlore-states}(c)}.
\end{array} \right.
\end{equation}
The energy is lowest in the state with equal link amplitudes that is related
to the classical state with equal angles between the spins
[Fig.~\ref{fig-pyrochlore-states}(c)].  We have evaluated the energy
difference between the states (a) and (c) analytically and found that $
(E_{\rm (a)} - E_{\rm (c)})/(JN\mathcal N) = \kappa^3/72$.
Note that the splitting is rather small: for the nominal equivalent of spin
1/2 ($\kappa=1$), the energies per spin differ by about one percent of the
exchange constant.  The order-from-disorder effect is extremely weak.

Unfortunately, we have been unable to prove that the state described
above is indeed the state of lowest energy, for we have discovered
another state with the same energy at $\mathcal O(\kappa^3)$.  Without
going into details, we note that the other state has a larger unit
cell and contains tetrahedra with ``coplanar'' spins
[Fig.~\ref{fig-pyrochlore-states}(b)].  

{\bf Outlook}. We have learned that finding {\em the} ground state of 
the Sp($N$) pyrochlore antiferromagnet is a hard problem.  For all 
previously studied systems, our method yields unique ground 
states in agreement with numerical minimizations at lowest non-trivial
order in $\kappa$.  
In contrast, simplices of the pyrochlore have continuously degenerate
ground states.  Apart from flux expulsion, there is no simple
principle that can guide the search for the ground state in this case.

At the same time, our study has produced some useful insights.  
First and foremost, we find that there is no ground state
retaining the full symmetry of the Hamiltonian; such a state is
already ruled out at the level of a single tetrahedron, in a
controlled fashion at small $\kappa$. {\em In $\spn$, there is no spin
liquid on the pyrochlore lattice at zero temperature}. Furthermore, there
is a huge number of nearly degenerate saddle points, which are not
related by a symmetry, with a splitting $\mathcal O(\kappa^3)$.
These small splittings suggest that determining the precise nature of
the symmetry breaking is going to be very hard and, for
experimental systems, exquisitely sensitive to small additional terms
in the Hamiltonian. 

In the large-$N$ treatment, tiny energy differences between saddle
points are made infinite as they come with a large prefactor
$N\rightarrow\infty$. In practice, $N = 1$, and therefore local 
tunneling events will
probably play a role. Whereas individual saddle points do break
symmetries, tunneling can lead to symmetry restoration.  At the
end of our investigation, we therefore have to declare ourselves
agnostic as to the eventual fate of the quantum pyrochlore magnet at
zero temperature. Whatever order may be present there will 
likely melt rapidly at finite temperature.

{\bf Acknowledgements.}  We are grateful to Y.B. Kim, B. Marston, and
S.  Sachdev for useful discussions.  O.T. acknowledges hospitality of
Ecole Polytechnique F\'ed\'erale de Lausanne.  This work was supported
in part by the Minist\`ere de la Recherche et des Nouvelles
Technologies, by the NSF Under Grants No. DMR-9978074, 0213706,
0348679, PHY99-07949, and by the David and Lucile Packard Foundation.


\begin{thebibliography}{99}

\bibitem{greedan} J.E.~Greedan, J.~Mater.~Chem. {\bf 11}, 37 (2001); 
R. Moessner, Can. J. Phys. 79, 1283 (2001).

\bibitem{shender} E.F. Shender, Sov. Phys. JETP {\bf 56}, 178 (1982);
C. L. Henley, J. Appl. Phys. {\bf 61}, 3962 (1987).

\bibitem{ml2003} This work has recently been reviewed in 
G. Misguich and C. Lhuillier, cond-mat/0310405.

\bibitem{arovas88} D.P. Arovas and A. Auerbach, \prb {\bf 38}, 316 (1988); 
A. Auerbach and D. P. Arovas, \prl {\bf 61}, 617 (1988).

\bibitem{read90} N. Read and S. Sachdev, \prb {\bf 42}, 4568 (1990).

\bibitem{sachdev92} S. Sachdev, in 
{\em Low-dimensional quantum field theories for condensed-matter
physicists}, ed. by Y. Lu, S. Lundqvist, and G. Morandi, World
Scientific, Singapore (1995); cond-mat/9303014.

\bibitem{sachdev92prb} S. Sachdev, \prb {\bf 45}, 12377 (1992).

\bibitem{fn-flux} This flux is, of course, central to the 
discussion of flucutations {\it about} the mean-field vacuum, as
in Ref.~\onlinecite{read90},
but its utility in organizing the mean-field theory {\em itself} has been
overlooked.

\bibitem{ts02} O.~Tchernyshyov and S.L.~Sondhi, Nucl. Phys. B {\bf 639},
429 (2002).

\bibitem{tmslong} O.~Tchernyshyov, R.~Moessner and S.L.~Sondhi,
(in preparation).

\bibitem{hbb} A.B.~Harris, A.J.~Berlinsky, and C.~Bruder,
J. Appl. Phys. {\bf 69}, 5200 (1991).

\bibitem{tch02} O.~Tchernyshyov, R.~Moessner, and S.L.~Sondhi,
\prb {\bf 66}, 064403 (2002).

\bibitem{tsune} H.~Tsunetsugu, \prb {\bf 65}, 024415 (2002).

\end{thebibliography}
\end{document}